# Ultrafast co-electrophoretic fluorescent staining of proteins with carbocyanines.

Sylvie Luche [1,2], Cécile Lelong [1,2], Hélène Diemer [3], Alain Van Dorsselaer [3], Thierry Rabilloud [1,2]

[1] CEA- Laboratoire de Biochimie et Biophysique des systèmes Intégrés, iRTSV/BBSI, CEA-Grenoble, 17 rue des martyrs, F-38054 GRENOBLE CEDEX 9, France

2 CNRS UMR 5092, CEA-Grenoble, 17 rue des martyrs, F-38054 GRENOBLE CEDEX 9, France

3 Laboratoire de Spectrométrie de Masse Bio-Organique, UMR CNRS 7178, ECPM, 25 rue Becquerel, 67087 STRASBOURG Cedex2, France

Correspondence :
Thierry Rabilloud, iRTSV/BBSI, UMR CNRS 5092,
CEA-Grenoble, 17 rue des martyrs,
F-38054 GRENOBLE CEDEX 9
Tel (33)-4-38-78-32-12
Fax (33)-4-38-78-44-99

e-mail: Thierry.Rabilloud@ cea.fr

abstract

Abstract

Protein detection on SDS gels or on two dimensional gels must combine several features, such as sensitivity, homogeneity from one protein to another, speed, low cost and user-friendliness. For some applications, it is also interesting to have a non fixing stain, so that proteins can be mobilized from the gel for further use (electroelution, blotting).
We show here that co-electrophoretic staining by fluorophores of the oxacarbocyanine family, and especially diheptyloxacarbocyanine, offers several positive features. The sensitivity is intermediate between the one of colloidal Coomassie blue and the one of fluroescent ruthenium complexes. Detection is achieved within one hour after the end of the electrophoretic process and does not use any fixing or toxic agent. The fluorescent SDS-carbocyanine-protein complexes can be detected either with a laser scanner with an excitation wavelength of 488nm or with a UV table operating at 302nm. Excellent sequence coverage in subsequent mass spectrometry analysis of proteolytic peptides is also achieved with this detection method.

1. Introduction

The analysis of proteins separated on electrophoresis gels, mono- or bidimensional, relies on the ability to detect them in a suitable way. The ideal protein detection protocol should be sensitive, homogeneous from one protein to another and linear throughout a wide dynamic range. In addition, suitable features include compatibility with digestion and mass spectrometry, speed, convenience and low cost.
Up to now, no protein detection method matches perfectly these prerequisites. Colloidal Coomassie Blue [1] is simple, cheap and rather linear, but its lacks sensitivity and requires long staining times for optimal sensitivity. Conversely, silver staining is sensitive, but labor-intensive, and its linearity is limited. Moreover, despite recent advances [2], [3], its compatibility is limited. Fluorescent detection methods, on their side, show a convenient linearity and an adequate sensitivity, although the latter varies from the one of colloidal Coomassie [4] to the one of silver [5], [6] . Although superior to the one of silver staining, their compatibility with mass spectrometry does not always equal the one of Coomassie Blue, and this has been unfortunately shown to be the case for Sypro Ruby [7] and Deep Purple [8], i.e. the most sensitive variants.
In addition to this drawback, optimal sensitivity requires rather long staining times, and the cost of the commercial reagents can become a concern when large series of gels are to be produced.

Last but not least, all these staining methods take place along with gel fixation, i.e. protein insolubilization in the gel. This means in turn that recovery of proteins after staining, e.g. for electroelution or blotting purposes, is rather problematic.
There are however a few exceptions to the latter rule. Protein staining with zinc and imidazole [9] is rather sensitive, but absolutely not linear. In addition, protein excision for mass spectrometry is difficult because of poor contrast.
Fluorescent staining with Nile Red [10] is rapid and does not require fixation. However, the staining is quite photolabile, which complicates imaging and excision, and the sensitivity is limited.
Finally, staining with Sypro Tangerine [11] just requires a single bath in saline solution containing the fluorophore. However, its cost and limited sensitivity have limited its use.

So the situation is worse in this category of non-fixing detection than in the case of classical detection after fixation. We report here protein detection with carbocyanines in SDS gels. This method is most efficient by co-electrophoresis of proteins with the fluorophore. This provides in turn excellent staining speed (30 minutes between end of electrophoresis and imaging). In addition, the method is cost-effective and compatible with several downstream processes, e.g. mass spectrometry or blotting. Finally, the sensitivity is intermediate between the one of Coomassie blue and the one of ruthenium complexes.

2. Material and methods

2.1. Samples

Molecular weight markers (broad range, Bio-Rad) were diluted down to 10 ng/µl for each band in SDS buffer (Tris-HCl 125mM pH 7.5, containing 2% (w/v) SDS, 5% (v/v) thioglycerol, 20% (v/v) glycerol and 0.005% (w/v) bromophenol blue). The diluted solution was heated in boiling water for 5 minutes. A tenfold dilution in SDS buffer was performed to get a 1ng/µl per protein dilution

JM 109 E. coli cells were grown up to an OD of 0.5 in LB medium. Bacteria were collected by centrifugation (2000g 5 minutes) and washed twice in PBS, and once in isotonic wash buffer (Tris-HCl 10mM pH 7.5, 0.25M sucrose, 1mM EDTA). The final pellet was suspended in its volume of isotonic wash buffer, transferred in an ultracentrifuge tube, and 4 volumes of concentrated lysis solution (8.75M urea, 2.5M thiourea, 5% CHAPS, 50mM DTT and 25mM spermine base) were added. After lysis at room temperature for 30 minutes, the viscous lystae was centrifuged at 200,000g for 1 hour at room temperature. The supernatant was collected, the protein concentration was estimated and the solution was made 0.4% (w/v) in carrier ampholytes (Pharmalyte 3-10). The solution was stored frozen at -20°C until use.

J774 cells and HeLa cells were grown in spinner flasks in DMEM + 5% fetal calf serum up to a density of 1 million cells /ml. The cells were collected by centrifugation (1000g 5 minutes), washed and lysed as described for bacteria.

2.2 Electrophoresis

2.2.1. SDS electrophoresis

10%T gels (160x200x1.5 mm) were used for protein separation. The Tris taurine buffer system was used [12], operated at a ionic strength of 0.1 and a pH of 7.9. The final gel composition is thus Tris 180mM, HCl 100 mM, acrylamide 10% (w/v), bisacrylamide 0.27%. The upper electrode buffer is Tris 50mM, Taurine 200mM, SDS 0.1%. The lower electrode buffer is Tris 50mM, glycine 200mM, SDS 0.1%.

For 1D SDS gels, a 4% stacking gel in Tris 125mM, HCl 100mM was used. No stacking gel was used for 2D electrophoresis.

The gels were run at 25V for 1hour, then 12.5W per gel until teh dye front has reached the bottom of the gel.

Alternatively, the standard Tris-glycine system, operating at pH 8.8 and ionic strength of 0.0625 was used.

2.2.2. IEF
Home made 160mm long 4-8 or 3-10.5 linear pH gradient gels were cast according to published procedures [13]. Four mm-wide strips were cut, and rehydrated overnight with the sample, diluted in a final volume of 0.6ml of rehydration solution (7M urea, 2M thiourea, 4% CHAPS and 100mM dithiodiethanol [14], [15]).
The strips were then placed in a multiphor plate, and IEF was carried out with the following electrical parameters

100V for 1 hour, then 300V for 3 hours, then 1000V for 1 hour, then 3400 V up to 60-70 kVh.

After IEF, the gels were equilibrated for 20 minutes in Tris 125mM, HCl 100mM, SDS 2.5%, glycerol 30% and urea 6M. They were then transferred on top of the SDS gels and sealed in place with 1% agarose dissolved in Tris 125mM, HCl 100mM, SDS 0.4% and 0.005% (w/v) bromophenol blue. Electrophoresis was carried out as described above.

2.3. Detection on gels

Colloidal coomassie blue staining was performed using a commercial product (G-Biosciences) purchased from Agro-Bio (La Ferté Saint Aubin, France).

Silver staining was performed according to the fast silver staining method of Rabilloud [16].

Ruthenium fluorescent staining was performed according to Lamanda et al [17], and Sypro tangerine staining was performed in PBS accoding to Steinberg et al [11]. The gels were imaged with a Fluorimager laser scanner, operating at 488nm excitation wavelength.

For co-electrophoretic carbocyanine staining, the desired carbocyanine (all were purchased from Fluka, Switzerland) was first dissolved at 30mM in DMSO.This stock solution is stable for severla weeks at room temperature, provided it is stored in the dark.
Various carbocyanines were tested, varying in their aromatic nucleus (oxacarbocyanine, thiacarbocyanine and tertramethylindocarbocyanine) and in the size of the side chains (ethyl to octadecyl). All carbocyanines can be excited at 302nm on a UV table. Oxacarbocyanines are also excited at 488nm, while thia- and indo-carbocyanines can be excited at 532nm.
The carbocyanine of interest was dissolved at 3μM (final concentration) in the upper electrode buffer. The fluorescent electrode buffer was used in place of standard buffer.
After electrophoresis, the colored gel was rinsed with 5-10 volumes of water. One 15 minutes rinse is sufficient for Tris-glycine gels operating at an ionic strength of 0.0625. Two 15 minutes rinses are required for gels operating at an ionic strength of 0.1. The gels were then visualized on a UV table or using a laser scanner.
Alternate washing solutions (e.g. 5% acetic acid, or 5% acetic acid + 20% ethanol, or 0.05% SDS) were also tested. they often required longer washing times to provide contrast, without leading to increased sensitivity.
For staining decay experiments, the gels were stained as described above, using water as a contrasting agent. After the initial rinses, the gels were scanned, and then returned to their water bath without shaking. The gels were then scanned at hourly intervals, and retruned to their bath between each scan

For post-electrophoretic carbocyanine staining, the general scheme described by Malone [18] was used and adapted as follows/
After electrophoresis, the gels are fixed overnight in 2% phosphoric acid, 30% ethanol and 0.004% SDS. They are then washed 3 x 20 minutes in 2% phosphoric acid, 0.004% SDS, and then equilibrated for 4 hours in 2% phosphoric acid, 0.004% SDS containing 3 μM of the desired carbocyanine. Visualization could be carried out directly from the carbocyanine bath on a UV table or using a laser scanner.

2.4. Image analysis
Images acquired on a Fluorimager laser scanner (488nm excitation wavelength, no emission filter, 200 micron pixe size) were analyzed directly by the ImageQuant software provided with the instrument for SDS-PAGE gels of markers. For two-dimensional gels, the images were converted to the TIFF format, and then analyzed with the delta 2D sotware (Decodon, Germany). The default detection parameters calculated by the software were used and no manual edition of the spots was performed.

2.5. Blotting
Proteins were transferred on PVDF membranes using the semi-dry blotting system of Kyhse-Andersen [19]. Gels can be transferred directly after electrophoresis. Otherwise, gels stained with carbocyanines were first re-equilibrated for 30 minutes in carbocyanine-free electrode buffer prior to transfer. Mock-colored gels, i.e. gels run with a colorless-electrode buffer but rinsed 2x15 minutes in water, were also reequilibrated prior to transfer.
After transfer, the proteins were detected on the PVDF sheets with india ink [20].

2.6. Mass spectrometry

2.6.1. Spot excision:
For fluorescent stains, spot excision was performed on a UV table operating at 302nm. The spots were collected in microtiter plates. The spots were generally not destained prior to acetonitrile washing. However, the carbocyanine-stained spots were sometimes fixed in 50% ethanol for 30 minutes. The solvent was then removed and the spots were stored at -20°C until use.

2.6.2. In gel digestion :
In gel digestion was performed with an automated protein digestion system, MassPrep Station
(Waters, Manchester, UK). The gel plugs were washed twice with 50 μL of 25 mM ammonium hydrogen carbonate (NH4HCO3) and 50 μL of acetonitrile. The cysteine residue were reduced by 50 μL of 10 mM dithiothreitol at 57°C and alkylated by 50 μL of 55 mM iodoacetamide. After dehydration with acetonitrile, the proteins were cleaved in gel with 10 μL of 12.5 ng/μL of modified porcine trypsin (Promega, Madison, WI, USA) in 25 mM NH4HCO3. The digestion was performed overnight at room temperature. The generated peptides were extracted with 60% acetonitrile in 5% acid formic.

2.6.3. MALDI-TOF-MS analysis
MALDI-TOF mass measurements were carried out on UltraflexTM TOF/TOF (Bruker DaltonikGmbH, Bremen, Germany). This instrument was used at a maximum accelerating potential of 25kV in positive mode and was operated in reflectron mode. The sample were prepared by standard dried droplet preparation on stainless steel MALDI targets using α-cyano-4-hydroxycinnamic acid as matrix.
The external calibration of MALDI mass spectra was carried out using singly charged monoisotopic peaks of a mixture of bradykinin 1-7 (m/z=757.400), human angiotensin II (m/z=1046.542), human angiotensin I (m/z=1296.685), substance P (m/z=1347.735), bombesin (m/z=1619.822), renin (m/z=1758.933), ACTH 1-17 (m/z=2093.087) and ACTH 18-39 (m/z=2465.199). To achieve mass accuracy, internal calibration was performed with tryptic peptides coming from autolysis of trypsin, with respectively monoisotopic masses at m/z = 842.510, m/z = 1045.564 and m/z = 2211.105 . Monoisotopic peptide masses were automatically annotated using Flexanalysis 2.0 software.

Peaks are automatically collected with a signal to noise ratio above 4 and a peak quality index greater than 30.

2.6.4. MS Data analysis
Monoisotopic peptide masses were assigned and used for databases searches using the search engine MASCOT (Matrix Science, London, UK) [21]. All proteins present in Swiss-Prot were used without any pI and Mr restrictions. The peptide mass error was limited to 50 ppm, one possible missed cleavage was accepted.

3. Results and discussion

Three different principles can be used for fluorescent detection on gels. The most obvious one, covalent labelling, has been used with limited sensitivity in the past with UV-excitable probes such as fluorescamine [22] or MDPF [23]. An important improvement of this approach has been made recently with the introduction of multiplex labelling [24]. However, the low number of labelling sites on proteins limits the absolute sensitivity of this approach. While this can be partly compensated by the increase of software performance for pure detection purposes, the approach suffers from a weak signal, which renders spot excision for characterization purposes difficult.
The second obvious approach is to use a probe that binds non covalently to proteins, in which case the contrast between the fluorescent protein zones and the background is governed by the differential concentration of the probes in the two sites. This approach is used in the very popular staining methods with ruthenium complexes [5] [17] or with epinocconone [6].
The last approach uses molecules which fluorescence depends on the chemical environment in which they are. Classical examples are Nile red [10], Sypro Orange and Sypro Red [4] and Sypro Tangerine [11]. This latter approach suffers generally from a lack of sensitivity. Despite commercial claims, their sensitivity is close to the one of colloidal Coomassie Blue for a substantially higher cost. This is not true for Nile red, but the limited solubility of this dye and its photolability limit the practicability of this approach.
As a matter of facts, the ideal fluorescent probe in this family should combine a high absorptivity, an important differential quantum yield between hydrophilic and lipophilic environment, yet being water-soluble. Last but not least, the molecule should have a high photostability to provide flexibility to the staining, e.g. for spot excision.
Carboyanines are a family of molecules that show many of these positive features. They are indeed at the basis of the high sensitivity multiplex labelling methods [24].
However, initial trials of incubating electrophoresis gels in dilute carbocyanine solutions, mimicking the Sypro Orange protocol, did not give any interesting result. We therefore tried to infuse the dye in the gel by co-electrophoresis along with the proteins, as has been described with Coomassie Blue [25]. Interesting results were obtained using water or water-alcohol-acid mixtures as the contrast-developing agent after electrophoresis, as shown in figure 1. However, acidic fixatives showed a tendency to develop an intense background in the low molecular weight region of the gels when carrier ampholytes are present, as in the case of 2D electrophoresis. Furthermore, the suggestion to lower the SDS concentration [25] induces some tailing in the 2D spots, as shown in figure 2.

These results prompted us to carry out a more thorough investigation of the structure-efficiency relationships in the carbocyanine family.

Carbocyanine can vary in the side chains and/or in the aromatic nucleus, as shown on figure 3. As the staining is supposed to be driven by the interaction between the probe and the protein-SDS complexes, a long side chain is supposed to increase the affinity for lipophilic environments and thus a strong fluorescence. However, a long side chain also increases the likelihood of self-aggregation of the probe (inducing background) and also decreases the solubility of the probe. The results of such tests are shown in figure 4. This figure clearly shows an optimum at moderately long side chains. Too short side chains (ethyl) show poor sensitivity, while very long side chains (octadecyl) are not very soluble and cannot be used at optimal concentrations, thereby limiting sensitivity.

Initial tests were carried out at 3μM carbocyanine concentration, i.e. 1 molecule of probe per 1000 molecules of SDS. While lowering this concentration decreased the sensitivity of staining, increasing it did not enhance staining. We therefore kept the 3μM concentration, which also allowed a very economical staining.

We also tried to replace the oxacarbocyanines with indocarbocyanines or with thiacarbocyanines. Tests on a UV table, the only "universal" source able to excitate all these molecules, suggested that oxacarbocyanines were the optimal family to work with (data not shown).

At this point, we tried to determine the response factor of the staining for different proteins. To this purpose, we used 1D electrophoresis of serial dilution of marker proteins, as shown in figure 1, and then quantified the fluorescence of the bands with the ImageQuant software. The results are shown on figure 5. The stain appears fairly linear for each protein, but diferent response factors are clearly obvious from one protein to another. Serum albumin showed the highest response factor, which might be linked to the propensity of this protein to bind various low molecular weight organic compounds.

We then compared this stain with the most closely stain operating under non-fixing conditions, i.e. Sypro Tangerine. The results are shown on figure 6. Sypro Tangerine is clearly inferior in sensitivity, especially in the low molecular weight region of the gel. This discrepancy between carbocyanine and Sypro tangerine could however be due to the diffusion of low molecular weight proteins in the gel during non-fixing staining, suggesting that the carbocyanine stain could be indeed a fixing one due to a special property of these molecules. To rule out this possibility, we performed blotting experiments, which are shown in figure 7. While a deleterious effect of the staining process can be noted, the protein can still be transferred efficiently to the membrane,while the stain went through the membrane.

Finally, we tested the sensitivity and the compatibility of this fluorescent stain with mass spectrometry, in comparison to standard methods such as colloidal Coomassie Blue [1] or ruthenium complexes [17]. The results are shown on figure 8 and on table 1. On the sensitivity level, the carbocyanine stain lies between Coomassie Blue and ruthenium complexes, while being much quicker to complete. A comparison of the homogeneity of staining was made between ruthenium, Coomassie Blue and carbocyanine. To this purpose, triplicate gels were run, stained and the resulting images were analyzed with the Delta2D software. The distributions of the staining intensities standard deviations are shown for both methods on figure 9.

On the mass spectrometry compatibility level, the carbocyanine stain was comparable to colloidal Coomassie Blue and to ruthenium complexes,as shown on Table 1 on spots showing detected with the three staining methods but probing a range of staining intensities, and spread on various positions of the gels. Due to the non-fixing nature of the stain, we could use the spots as coming from the gels, i.e. without fixation, but with the buffer components. In this case, the usual spot washes prior to digestion were the only cleaning step. Alternatively, we could also fix the excised spots with classical acid-alcohol mixtures, thereby insolubilizing the proteins more efficiently in the gel but also providing more extensive cleaning by removal of electrophoresis buffer components.
Finally, it can be noted that carbocyanines offer excellent resistance to photoinduced fading, allowing ample time for spot excision on a UV table.

Quite differently from the method described in patents [26], this cyanine staining method is based on the interaction of oxacarbocyanine molecules with SDS-protein complexes, and is thus non-fixing. As the cyanine fluorescence is maximal in lipophilic environments, the rationale of this method is to disrupt SDS micelles causing the background fluorescence, while keeping enough SDS at the protein containing sites to promote fluorescence. This is achieved by water rinses, which decrease the ionic strength of the gel, thereby increasing the critical micellar concentration of SDS [27], and thus promoting SDS micelles dissociation. This means in turn that the protocol has to be adapted to the electrophoretic system used, as shown in figure 10. While a single rinse is optimal for systems operating at low ionic strength, such as the popular Laemmli system, increased rinses are required for systems operating at higher ionic strength, such as the Tris taurine system.
While a non-fixing process provides interesting features (speed of staining, further mobilization of the proteins, e.g. for protein blotting), it also means in turn that the stain decreases and the proteins diffuse with time, as shown in figure 11. Evaluation of the total stain by the Delta 2D software showed a 10% decrease in total signal intensity over 6 hours, but it can easily be seen that the fuzziness of the stain increases. Another staining scheme can also be devised, starting from gels that are fixed in the presence of SDS, as described by Malone et al. [18]. Typical results are shown on figure 12. In this case, the short chain oxacarbocyanines proved more efficient than carbocyanines with longer chains.

4. Concluding remarks

Carbocyanine staining is based on the increase of fluorescence of these molecules in the SDS-protein complexes obtained after SDS PAGE. Two setups can be used. The long setup, starting from fixed gels, gives a stable staining pattern. However, it is not as sensitive as ruthenium complexes, almost as long, so that its only advantage is economical.
The short setup, which uses coelectrophoresis of the fluorescent probe along with the proteins in the SDS gel, does not give a stable pattern over time, so that its applicability for very large gel series is limited. However, the staining pattern is stable for more than one hour, which allows ample time for image acquisition.
In addition this short setup provides a very fast (less than one hour), economical, sensitive (intermediate between Coomassie Blue and ruthenium complexes) and versatile stain because of its non fixative nature.

5. Acknowledgments.

The support from the Région Rhone Alpes by a grant (analytical chemistry subcall, priority research areas call 2003-2006) is gratefully acknowledged.

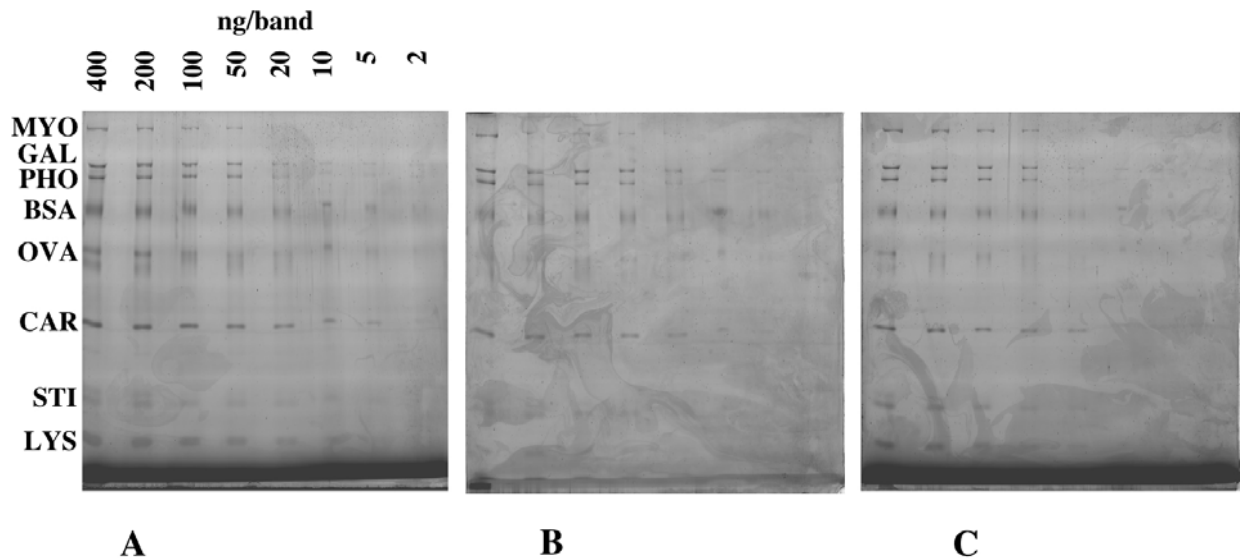

Figure 1: staining of molecular weight markers
serial dilutions of broad range molecular weight markers were loaded onto a SDS gel (10% acrylamide, Tris taurine system). the standard composition is the following: MYO: myosin; GAL: beta galatosidase; PHO: glycogen phosphorylase; BSA: bovine serum albumin; OVA: ovalbumin; CAR: carbonic anhydrase; STI: soybean trypsin inhibitor; LYS: lysozyme.
Loadings from left to right, per band of protein: 400ng, 200ng, 100ng, 50ng, 20ng, 10ng, 5ng, 2ng. The upper electrode buffer contained 3μM diheptyloxacarbocyanine and 0.1% SDS. After electrophoresis, the gels were rinsed 3x20 minutes prior to imaging
A: water rinses. B: rinses in 20% ethanol. C: rinses in 20% ethanol + 5% acetic acid

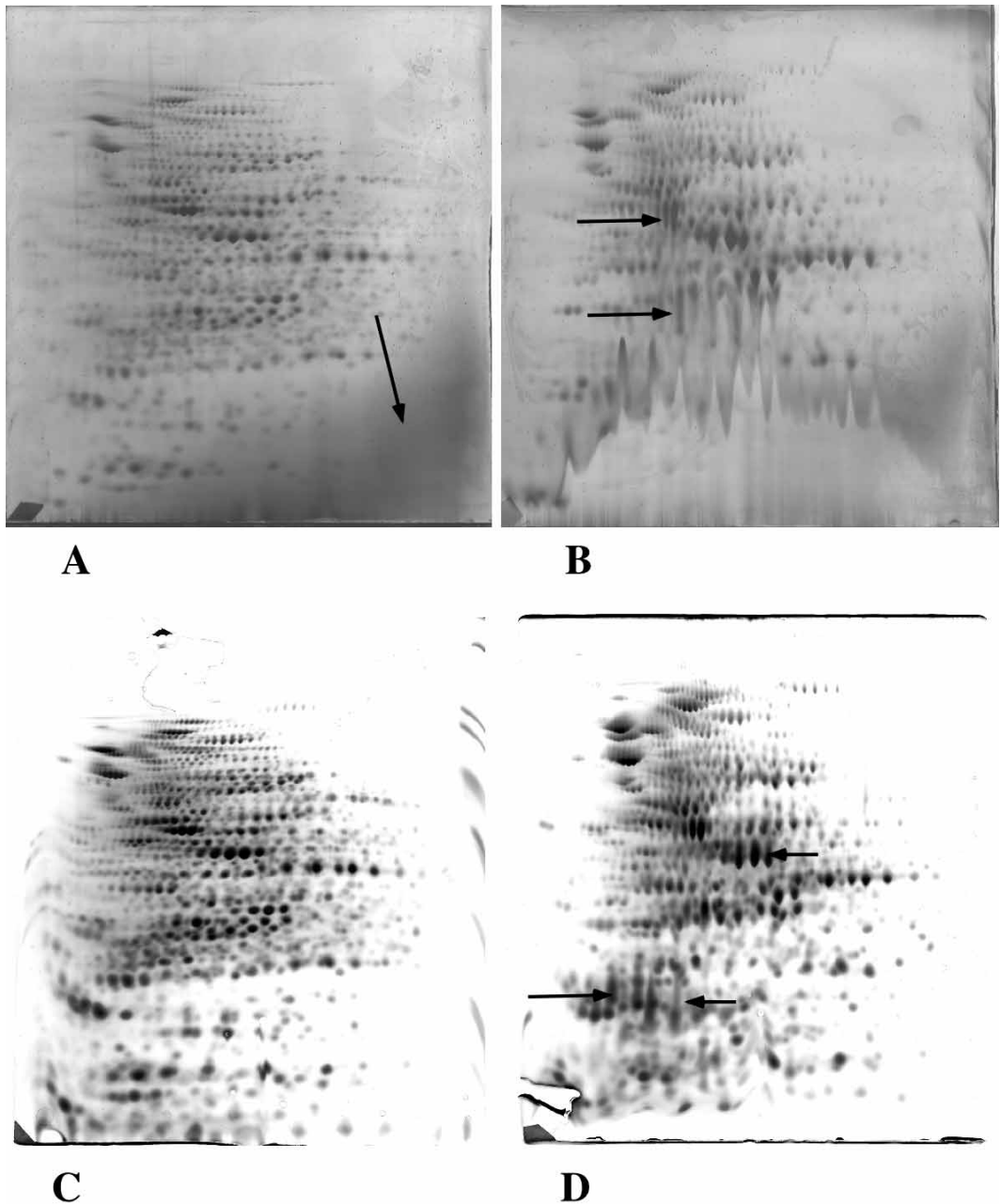

Figure 2: effect of SDS concentration
200 micrograms of E.coli proteins were separated by two dimensional electrophoresis (4-8 pH gradients, Tris taurine system). The electrode buffer contained 3µM diheptyloxacarbocyanine and 0.1% SDS (panel A) or 0.05% SDS (panel B). After electrophoresis, the gels were soaked 3x20 minutes in 5% acetic acid, 20% ethanol and 75% water (by volume) prior to imaging. Note the stained crescent of ampholytes (panel A, arrow) and the deformation of the lower part of the gels and the vertical tailing of some spots induced by the reduced SDS concentration (panel B, arrows). Ro check thatb this tailing is not just a staining artefact, a second pair of gels were run with normal (panel C) and reduced (panel D) SDS concentrations, without carbocyanine staining, and silver stained. Vertical tailing was also observed in this case

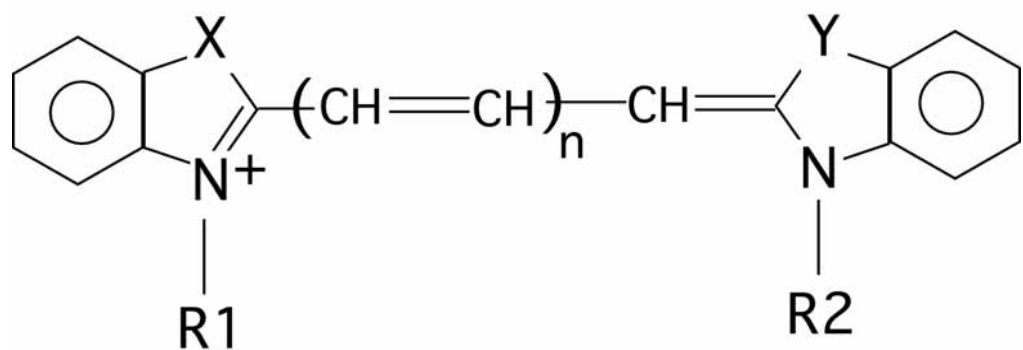

Figure 3: general structural formula of carbocyanines
carbocyanines can vary in the side chains (R1 and R2), in the bridge length (n) (n=1 in carbocyanines and n=2 in dicarbocyanines) and in the aromatic nucleus, where the X and Y positions can be occupied by an oxygen atom (oxacarbocyanines), a sulfur atom (thiacarbocyanines) or a dimethyl subsituted carbon atom (indocarbocyanines).

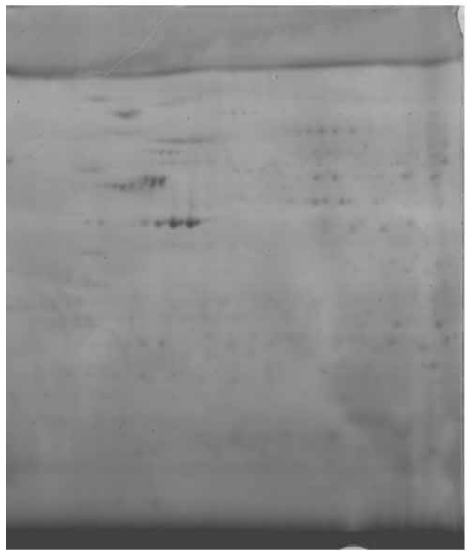 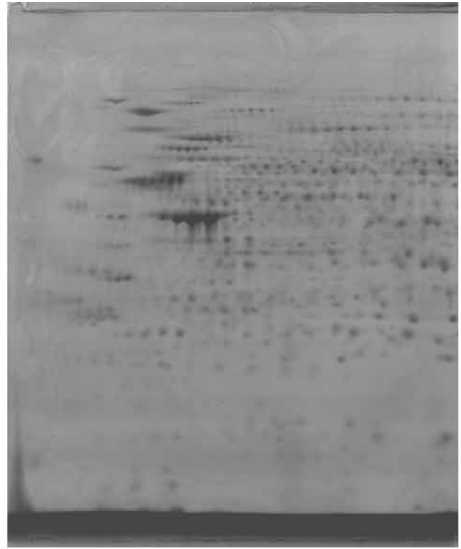
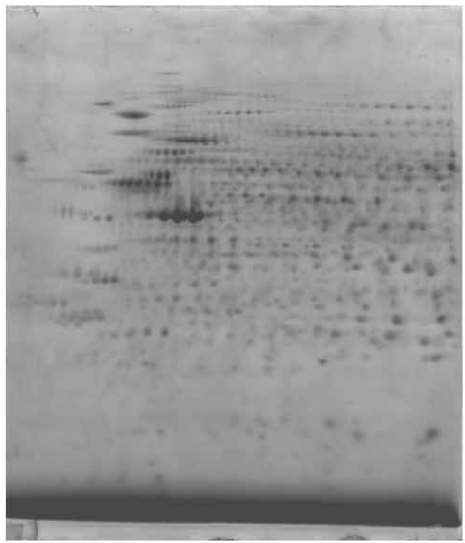 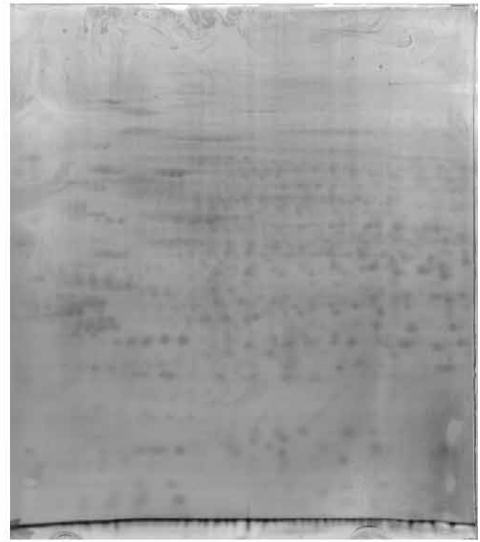

Figure 4: evaluation of optimal chain length in carbocyanine staining
200 micrograms of proteins (J774 cells) were separated by two dimensional electrophoresis (4-8 pH gradients, Tris taurine system). The gels were stained by co electrophoretic carbocyanine staining using diethyl oxacarbocyanine (panel A), dipentyl oxacarbocyanine (panel B), diheptyl oxacarbocyanine (panel C), dioctadecyl oxacarbocyanine (panel D). A sensitivity optimum appears at 5-7 carbons in the side chain.

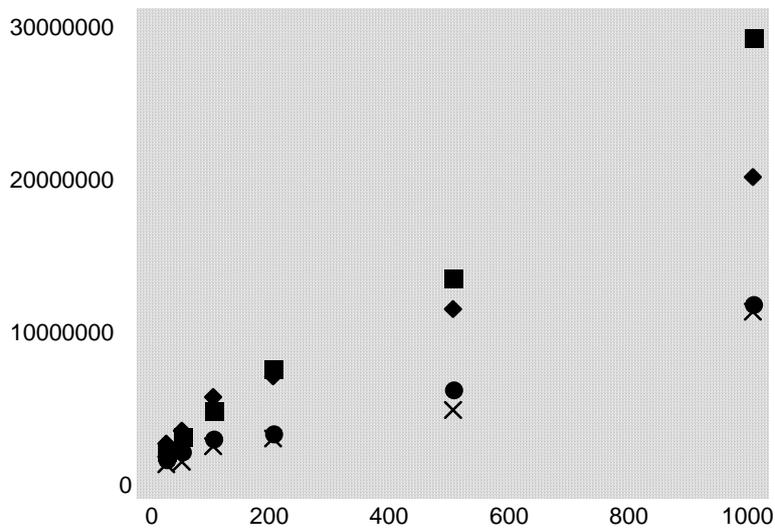

Figure 5 : Dynamic response factor for various proteins.
The molecular weight marker proteins used in figure 1 were separated by SDS PAGE, as detailed in figure 1. The gels were stained by co-electrophoretic staning with diheptyloxacarbocyanine, using water as the contrasting agent. The fluorescence intensity was then measured by the ImageQuant software, and plotted against protein load for different proteins. Squares : bovine serum albumin ; diamonds : hen ovalbumin ; circles : rabbit phosphorylase ; crosses : E. coli beta-galactosidase

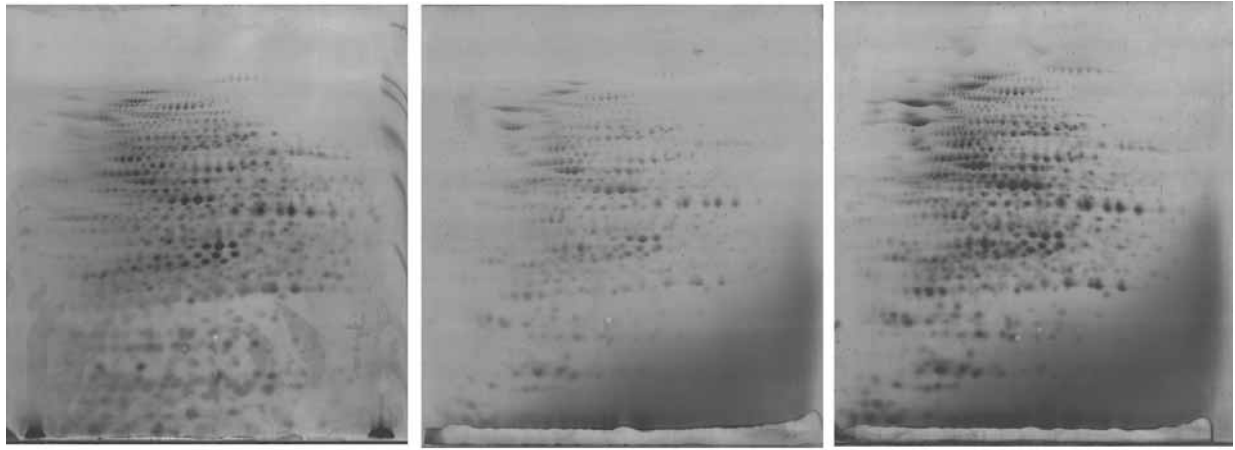

Figure 6: Comparison of carbocyanine staining with other commercial stains

200 micrograms of E.coli proteins were separated by two dimensional electrophoresis (4-8 pH gradients, Tris taurine system). The gels were stained by co electrophoretic carbocyanine staining (panel A), Sypro tangerine (panel B) or Sypro orange (panel C). Carbocyanine staining offers a better sensitivity than Sypro tangerine, which is the commercially available non-denaturing stain. The sensitivity is comparable to the one offered by Sypro Orange, which is slower to perform and immobilises proteins in th gel

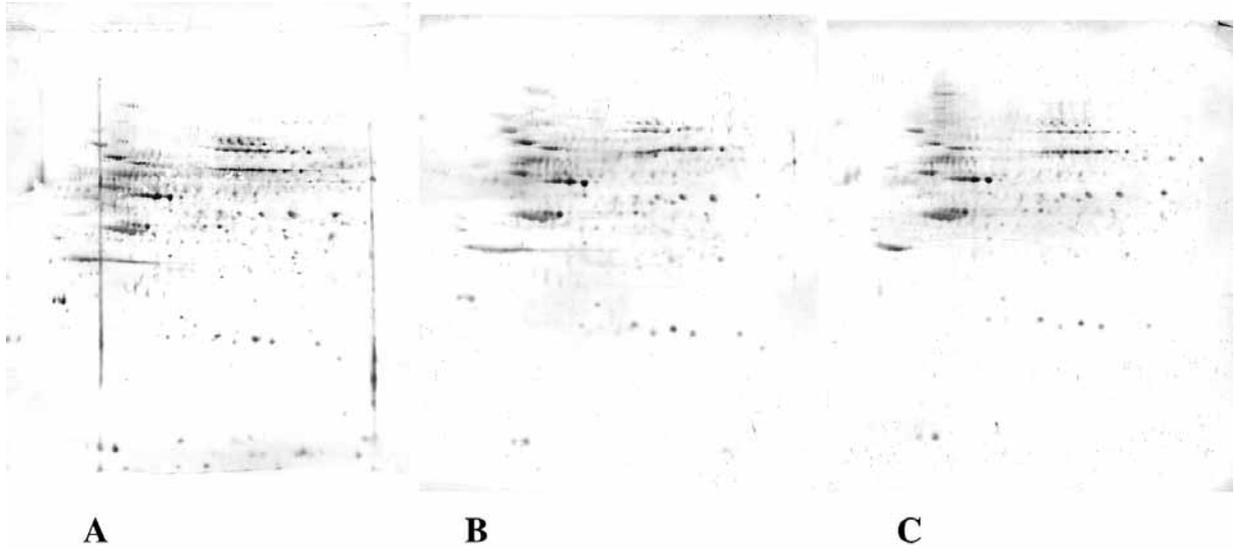

Figure 7: Blotting efficiency after carbocyanine staining

200 micrograms of proteins from a complete cell extract (HeLa cells) were separated by two dimensional gel electrophoresis (pH 4-8), using the Tris-Taurine system, either under control conditions (panels A and B) or with co-electrophoretic carbocyanine staining (panel C). After electrophoresis, the gel corresponding to panel A was blotted directly onto a PVDF membrane. The gels corresponding to panels B and C were rinsed 2x15 minutes in water, the reloaded in buffer by a 30 minutes soaking in Tris-taurine-SDS buffer (electrode buffer) prior to blotting. Protein transfer on the blot was detected by india ink staining

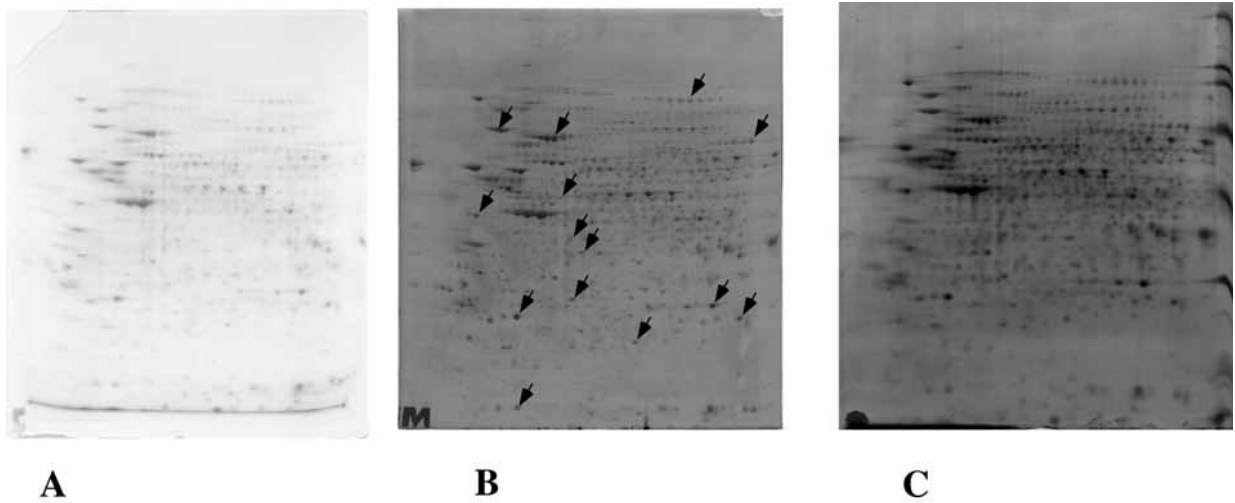

Figure 8: Comparison of staining by different methods.
300 micrograms (panel A) or 150 micrograms (panels B and C) of proteins from a complete cell extract (J774 cells) were separated by two dimensional gel electrophoresis (pH 3-10), using the Tris-Taurine system, either under control conditions (panels A and B) or with co-electrophoretic carbocyanine staining (panel C). The gel shown in panel A was then stained with Coomassie Blue with a commercial product and without prior fixation (note the intense blue zone corresponding to carrier ampholytes). The gel shown in panel B was stained with a fluorescent ruthenium complex. The spots excised for further characterization with mass spectrometry are shown with arrows

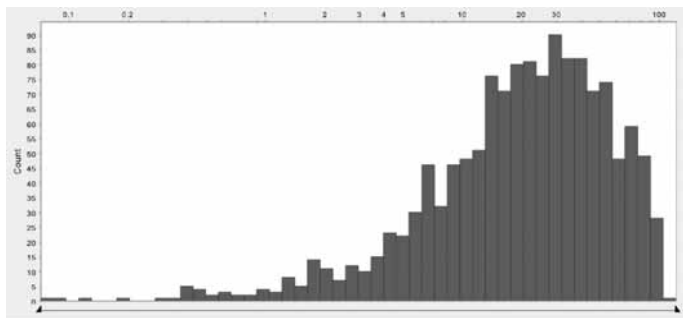

A

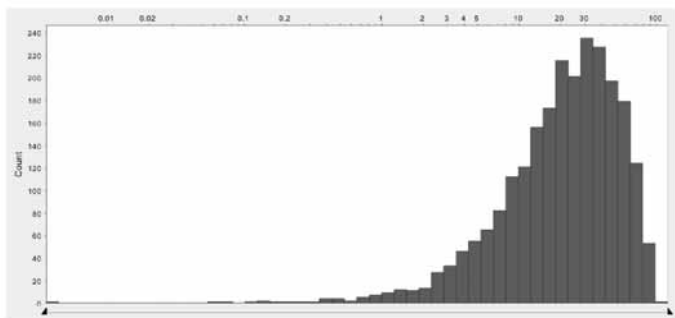

B

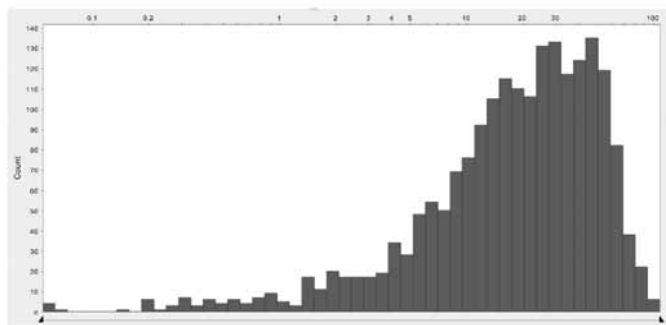

C

Figure 9 : relative standard deviation of spot quantitation with different detection methods. Triplicate gels were run according to the conditions and loadings detailed in figure 8. The spots were then quantitated and matched with the delta2D software, and the relative standard deviation (rsd) for quantitation (i.e. standard deviation/mean , experssed in percentile) was calculated for each standard method and displayed as a histogram.
A : histogram for carbocyanine staining (2571 spots detected)
B : histogram for ruthenium complex staining (3492 spots detected)
C : histogram for colloidal coomassie staining (2231 spots detected)
The median rsd (i.e. equal numbers of spots with a lower and higher rsd) are the following :
Carbocyanine : 19.6% ; Ruthenium : 20.1% ; Coomassie Blue 23.9%

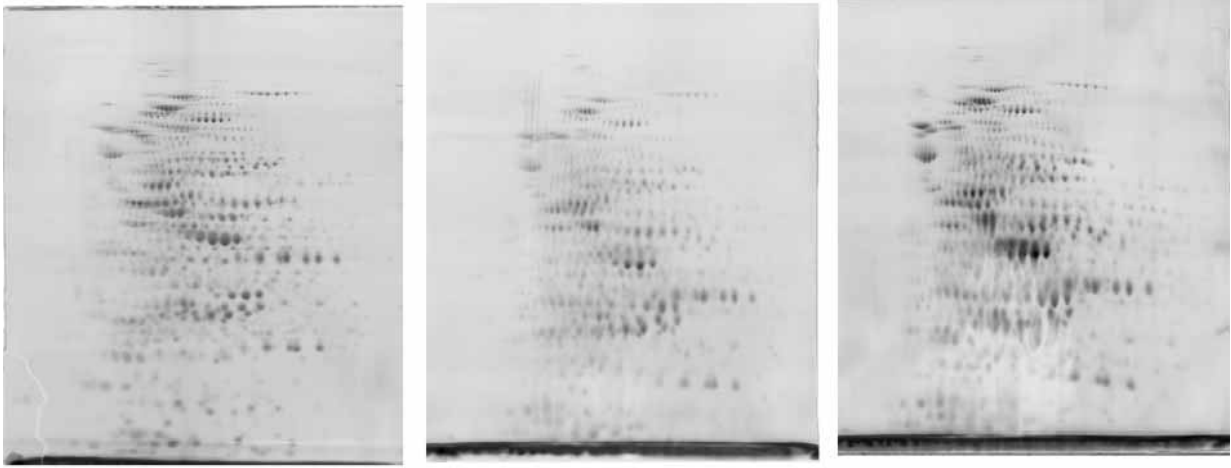

Figure 10: Effect of rinses in different buffer systems

200 micrograms of E.coli proteins were separated by two dimensional electrophoresis (4-8 pH gradients) and stained co-electrophoretically with carbocyanine. Gel A: Tris taurine system, operating at ionic strength 0.1, three water rinses of 15 minutes prior to imaging. Gel B: tris glycine system, operating at ionic strength 0.0625, three water rinses of 15 minutes prior to imaging. Gel C: tris glycine system, operating at ionic strength 0.0625, one water rinse of 15 minutes prior to imaging

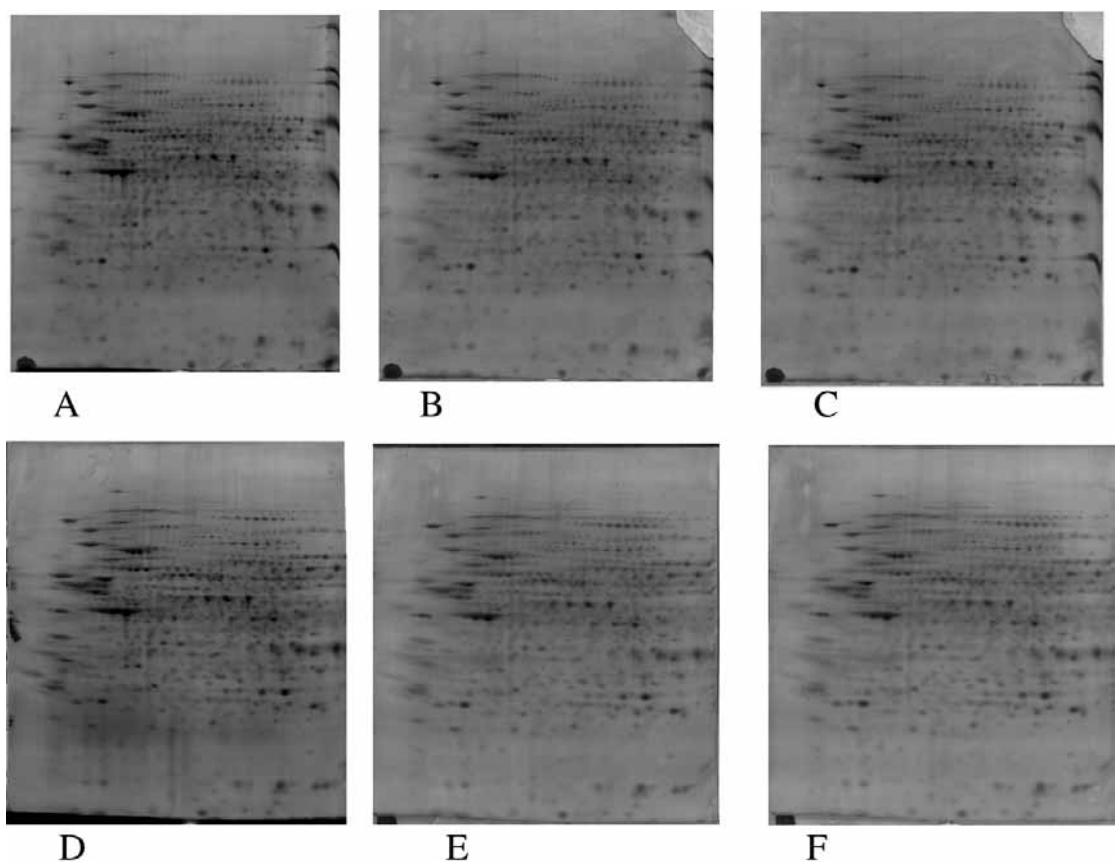

Figure 11: Stability of staining over time. 150 micrograms of proteins from a complete cell extract (J774 cells) were separated by two dimensional gel electrophoresis (4-8 pH gradients) and stained co-electrophoretically with carbocyanine. After the initial stain, the gels were left in water and scanned preiodically. Gel A: Starting gel (three water rinses, 15 minutes each). Gel B: two hours stay in water after first scan. Gel C: Three hours stay in water after gels scan.
Gel D: Starting gel (three water rinses, 15 minutes each). Gel E: four hours stay in water after first scan. Gel F: six hours stay in water after gels scan
Two sets of gels were required because of gel breakage upon repetitive manipulation.

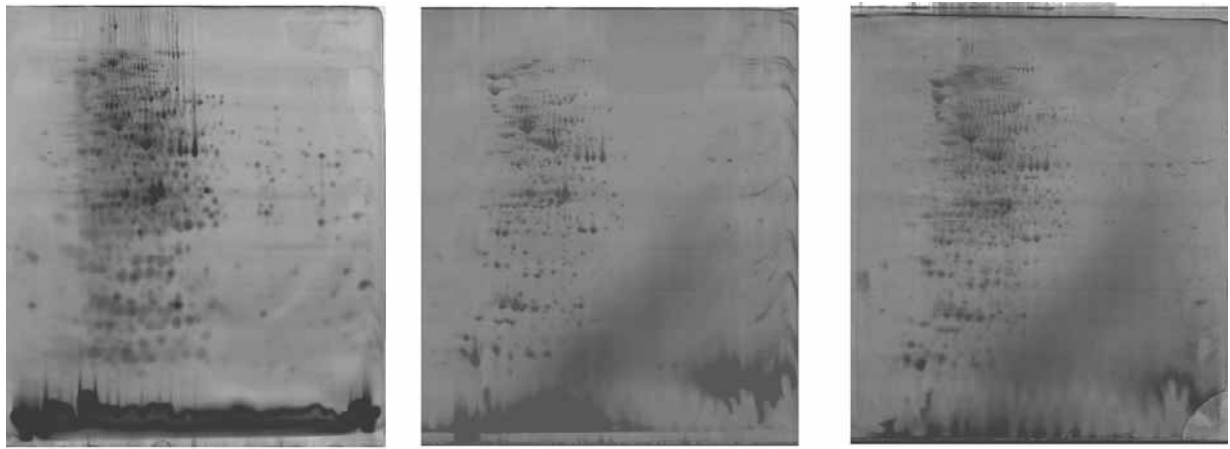

Figure 12: fixing staining with carbocyanines
300 micrograms of E.coli proteins were separated by two dimensional electrophoresis (3-10 pH gradients, Tris taurine system). The gels were stained co-electrophoretically with diheptyl oxacarbocyanine (panel A), or migrated without carbocyanine. The gels were then fixed and post-stained with dipentyl oxacarbocyanine (panel B) or diethyl oxacarbocyanine (panel C)

| STAIN | PROTEIN NAME | ACC. | MW / pI | %C | nb pep. |
|---|---|---|---|---|---|
| CBB | HEAT SHOCK PROTEIN 84B | Q71LX8 | 83229 / 4,57 | 56% | 43 |
| OF | HEAT SHOCK PROTEIN 84B | Q71LX8 | 83229 / 4,57 | 57% | 49 |
| ONF | HEAT SHOCK PROTEIN 84B | Q71LX8 | 83229 / 4,57 | 55% | 47 |
| Ru | HEAT SHOCK PROTEIN 84B | Q71LX8 | 83229 / 4,57 | 57% | 45 |
| CBB | ELONGATION FACTOR 2 | P58252 | 95122 / 6,42 | 25% | 21 |
| OF | ELONGATION FACTOR 2 | P58252 | 95122 / 6,42 | 34% | 29 |
| ONF | ELONGATION FACTOR 2 | P58252 | 95122 / 6,42 | 31% | 24 |
| Ru | ELONGATION FACTOR 2 | P58252 | 95122 / 6,42 | 54% | 41 |
| CBB | HEAT SHOCK COGNATE 71 KDA PROTEIN | P63017 | 70827 / 5,37 | 56% | 36 |
| OF | HEAT SHOCK COGNATE 71 KDA PROTEIN | P63017 | 70827 / 5,37 | 53% | 37 |
| ONF | HEAT SHOCK COGNATE 71 KDA PROTEIN | P63017 | 70827 / 5,37 | 63% | 44 |
| Ru | HEAT SHOCK COGNATE 71 KDA PROTEIN | P63017 | 70827 / 5,37 | 56% | 29 |
| CBB | TRANSKELOTASE | P40142 | 67588 / 7,23 | 50% | 32 |
| OF | TRANSKELOTASE | P40142 | 67588 / 7,23 | 56% | 23 |
| ONF | TRANSKELOTASE | P40142 | 67588 / 7,23 | 48% | 25 |
| Ru | TRANSKELOTASE | P40142 | 67588 / 7,23 | 64% | 39 |
| CBB | UBIQUINOL-CYTOCHROME-C REDUCTASE COMPLEX CORE PROTEIN I, … | Q9CZ13 | 52735 / 5,75 | 53% | 30 |
| OF | UBIQUINOL-CYTOCHROME-C REDUCTASE COMPLEX CORE PROTEIN I, … | Q9CZ13 | 52735 / 5,75 | 52% | 22 |
| ONF | UBIQUINOL-CYTOCHROME-C REDUCTASE COMPLEX CORE | Q9CZ13 | 52735 / 5,75 | 38% | 16 |

| | | | | |
|---|---|---|---|---|
| | PROTEIN I, … | | | |
| Ru | UBIQUINOL-CYTOCHROME-C REDUCTASE COMPLEX CORE PROTEIN I, … | Q9CZ13 | 52735 / 5,75 | 48% | 23 |
| CBB | FARNESYL PYROPHOSPHATE SYNTHETASE | Q92025 | 40556 / 5,46 | 35% | 13 |
| OF | FARNESYL PYROPHOSPHATE SYNTHETASE | Q920E5 | 40556 / 5,46 | 25% | 8 |
| ONF | FARNESYL PYROPHOSPHATE SYNTHETASE | Q920E5 | 40556 / 5,46 | 24% | 7 |
| Ru | FARNESYL PYROPHOSPHATE SYNTHETASE | Q920E5 | 40556 / 5,46 | 30% | 10 |
| CBB | ARBP PROTEIN (60S ACIDIC RIBOSOMAL PROTEIN P0) | Q5FWB6 | 34165 / 5,91 | 63% | 20 |
| OF | ARBP PROTEIN (60S ACIDIC RIBOSOMAL PROTEIN P0) | Q5FWB6 | 34165 / 5,91 | 45% | 13 |
| ONF | ARBP PROTEIN (60S ACIDIC RIBOSOMAL PROTEIN P0) | Q5FWB6 | 34165 / 5,91 | 49% | 12 |
| Ru | ARBP PROTEIN (60S ACIDIC RIBOSOMAL PROTEIN P0) | Q5FWB6 | 34165 / 5,91 | 55% | 15 |
| CBB | PHOSPHOGLYCERATE MUTASE 1 | Q9DBJ1 | 28683 / 6,75 | 80% | 24 |
| OF | PHOSPHOGLYCERATE MUTASE 1 | Q9DBJ1 | 28683 / 6,75 | 66% | 20 |
| ONF | PHOSPHOGLYCERATE MUTASE 1 | Q9DBJ1 | 28683 / 6,75 | 73% | 23 |
| Ru | PHOSPHOGLYCERATE MUTASE 1 | Q9DBJ1 | 28683 / 6,75 | 79% | 19 |
| CBB | GTP-BNDING NUCLEAR PROTEIN RAN | P62827 | 24277 / 7,19 | 66% | 21 |
| OF | GTP-BNDING NUCLEAR PROTEIN RAN | P62827 | 24277 / 7,19 | 65% | 17 |
| ONF | GTP-BNDING NUCLEAR PROTEIN RAN | P62827 | 24277 / 7,19 | 65% | 16 |
| Ru | GTP-BNDING NUCLEAR PROTEIN RAN | P62827 | 24277 / 7,19 | 61% | 18 |
| CBB | APRT PROTEIN (ADENINE PHOSPHORIBOSYL TRANSFERASE) | Q6PK77 | 19712 / 6,31 | 85% | 16 |
| OF | APRT PROTEIN (ADENINE PHOSPHORIBOSYL | Q6PK77 | 19712 / 6,31 | 66% | 12 |
| | TRANSFERASE) | | | | |
| ONF | APRT PROTEIN (ADENINE PHOSPHORIBOSYL TRANSFERASE) | Q6PK77 | 19712 / 6,31 | 68% | 11 |
| Ru | APRT PROTEIN (ADENINE PHOSPHORIBOSYL TRANSFERASE) | Q6PK77 | 19712 / 6,31 | 92% | 17 |
| CBB | EUKARYOTIC TRANSLATION INITIATION FACTOR 5A | P63242 | 16690 / 5,08 | 67% | 15 |
| OF | EUKARYOTIC TRANSLATION INITIATION FACTOR 5A | P63242 | 16690 / 5,08 | 64% | 7 |
| ONF | EUKARYOTIC TRANSLATION INITIATION FACTOR 5A | P63242 | 16690 / 5,08 | 67% | 12 |
| Ru | EUKARYOTIC TRANSLATION INITIATION FACTOR 5A | P63242 | 16690 / 5,08 | 70% | 12 |
| CBB | ANNEXIN A4 | P97429 | 35836 / 5,43 | 63% | 29 |
| OF | ANNEXIN A4 | P97429 | 35836 / 5,43 | 58% | 23 |
| ONF | ANNEXIN A4 | P97429 | 35836 / 5,43 | 63% | 24 |
| Ru | ANNEXIN A4 | P97429 | 35836 / 5,43 | 66% | 23 |
| CBB | RPSA PROTEIN (FRAGMENT N-TERMINAL) | Q58E74 | 32821 / 4,80 | 60% | 21 |
| OF | RPSA PROTEIN (FRAGMENT N-TERMINAL) | Q58E74 | 32821 / 4,80 | 56% | 15 |
| ONF | RPSA PROTEIN (FRAGMENT N-TERMINAL) | Q58E74 | 32821 / 4,80 | 56% | 16 |
| Ru | RPSA PROTEIN (FRAGMENT N-TERMINAL) | Q58E74 | 32821 / 4,80 | 61% | 19 |
| CBB | RHO GDP-DISSOCIATION INHIBITOR 1 | Q99PT1 | 23393 / 5,12 | 83% | 19 |
| OF | RHO GDP-DISSOCIATION INHIBITOR 1 | Q99PT1 | 23393 / 5,12 | 86% | 20 |
| ONF | RHO GDP-DISSOCIATION INHIBITOR 1 | Q99PT1 | 23393 / 5,12 | 87% | 19 |
| Ru | RHO GDP-DISSOCIATION INHIBITOR 1 | Q99PT1 | 23393 / 5,12 | 90% | 22 |

Table 1: MS analysis of proteins stained by different methods

Homologous spots excised from two-dimensional gels (4-8 linear pH gradients, 10% acrylamide) loaded with equal amounts of J774 proteins (200μg) and stained by various methods were digested, and the digests were analysed by MALDI mass spectrometry

First column: detection method: CBB= colloidal Coomassie Blue; OF: diheptyloxacarbocyanine, fixation of the spots post excision in aqueous alcohol; ONF: diheptyloxacarbocyanine, no fixation; Ru: ruthenium complex

Second and third column: protein name and SwissProt accession number, respectively

Fourth column: theoretical Mw and pI

Fifth column: sequence coverage of the MS analysis

Sixth column: number of observed peptides matching the protein sequence in the database